\documentclass[12pt]{article}
\usepackage{amssymb,amsmath,amscd}
\makeatletter

\@addtoreset{equation}{section}
\def\section{\@startsection {section}{1}{\z@}{-2.25ex plus -1ex minus
 -.2ex}{0.8ex plus .2ex}{\large\bf}}
\def\subsection{\@startsection{subsection}{2}{\z@}{-2.0ex plus%
 -1ex minus -.2ex}{0.4ex plus .2ex}{\bf}}

\def\isom{\phi_\tau}

\def\Ad{\mathrm{Ad}}
\def\ad{\mathrm{ad}}

\newcommand{\tr}[0]{{\rm tr}}

\newcommand{\inv}[0]{{-1}}

\newcommand{\cif}[0]{\mathcal{C}^\infty}

\newcommand{\cd}[0]{\!\cdot\!}

\newcommand{\al}{\alpha}
\newcommand{\bet}{\beta}

\def\bx{{\mbox{\boldmath $x$}}}

\def\by{{\mbox{\boldmath $y$}}}

\def\bp{{\mbox{\boldmath $p$}}}
\def\bP{{\mbox{\boldmath $P$}}}
\def\bq{{\mbox{\boldmath $q$}}}

\def\bJ{{\mbox{\boldmath $J$}}}
\def\bS{{\mbox{\boldmath $S$}}}
\def\bI{{\mbox{\boldmath $I$}}}
\def\bR{{\mbox{\boldmath $R$}}}
\def\bP{{\mbox{\boldmath $P$}}}
\def\bn{{\mbox{\boldmath $n$}}}

\def\tJ{\tilde J}
\def\btJ{\tilde{\mbox{\boldmath $J$}}}
\def\bm{{\mbox{\boldmath $m$}}}

\newcommand{\gothg}{\mathfrak g }

\newcommand{\gothh}{\mathfrak h }

\newcommand{\RR}{\mathbb{R}}
\newcommand{\CC}{\mathbb{C}}

\newtheorem{theorem}{Theorem}[section]
\newtheorem{lemma}[theorem]{Lemma}

\def\bea{\begin{eqnarray}}
\def\eea{\end{eqnarray}}
\def\bmz{\left(\begin{array}{2,2}}
\def\emz{\end{array}\right)}
\def\bmd{\left(\begin{array}{3,3}}
\def\emd{\end{array}\right)}

\newcommand{\tl}[0]{\theta}
\def\bpm{\begin{pmatrix}}
\def\epm{\end{pmatrix}}
\newcommand{\sqcc}{\sqrt{|\Lambda|}}

\begin{document}
\parskip 6pt
\parindent 0pt
\begin{flushright}
EMPG-08-05\\
pi-qg-79\\
\end{flushright}

\begin{center}
\baselineskip 24 pt {\Large \bf  Generalised Chern-Simons actions for 3d gravity and $\kappa$-Poincar\'e symmetry}

\baselineskip 16 pt

\vspace{.5cm} {{ C.~Meusburger}\footnote{\tt  cmeusburger@perimeterinstitute.ca}\\
Perimeter Institute for Theoretical Physics\\
31 Caroline Street North,
Waterloo, Ontario N2L 2Y5, Canada\\

\vspace{.5cm}
{ B.~J.~Schroers}\footnote{\tt bernd@ma.hw.ac.uk} \\
Department of Mathematics and Maxwell Institute for Mathematical Sciences \\
 Heriot-Watt University \\
Edinburgh EH14 4AS, United Kingdom } \\

\vspace{0.2cm}

{revised  version, 23 June   2008}

\end{center}

\begin{abstract}
\noindent
We consider Chern-Simons theories  for the 
Poincar\'e, de Sitter and anti-de Sitter groups in three dimensions
which generalise the Chern-Simons formulation of 3d gravity.
We determine  conditions under which  
 $\kappa$-Poincar\'e symmetry and its de Sitter and 
anti-de Sitter analogues  can be associated to these theories as 
 quantised symmetries. Assuming the usual form of 
those  symmetries, with a  timelike vector 
as deformation parameter,  we find that 
such an association is possible only in the de Sitter case, and that 
the associated  Chern-Simons action  is not the gravitational one.
Although  the resulting theory  and 
3d gravity have the same equations of motion for the gauge field,
they are not equivalent, even classically, since they differ in 
their  symplectic structure and  the coupling to matter.
We deduce that 
   $\kappa$-Poincar\'e symmetry is not 
 associated to either classical or quantum gravity in 
three dimensions. Starting from the (non-gravitational)
 Chern-Simons action
 we explain how to construct a multi-particle model which is 
invariant under the classical analogue of  $\kappa$-de Sitter symmetry,
and carry out the first steps in that construction. \end{abstract}
\centerline{PACS numbers: 04.20.Cv, 02.20.Qs, 02.40.-k}

\section{Introduction}
\subsection{Motivation}
The possibility of  deforming   Poincar\'e symmetry with a
dimensionful parameter (traditionally called $\kappa$) has attracted
much interest since its discovery seventeen years ago
\cite{LNRT,LNR}. The deformed symmetry, often called
$\kappa$-Poincar\'e algebra, is a Hopf algebra, whose mathematical
structure is now well understood \cite{MR}. However, both the role
of $\kappa$-Poincar\'e symmetry   as a fundamental symmetry
in physics and its phenomenological status remain unclear. It has been
argued that $\kappa$-Poincar\'e symmetry arises in a low-energy
limit of quantum gravity in four dimensions \cite{ACSS}.
However, since a quantum theory of gravity in four dimensions
could not be constructed so far, these arguments are largely
heuristic. At the same time, the derivation of experimentally
testable consequences from $\kappa$-Poincar\'e symmetry has been
hampered by the lack of   $\kappa$-Poincar\'e-invariant
theories with non-trivial interactions.

In this paper we address  both of the above
issues in three dimensions, where they turn out to be closely
 linked.  We consider a family of Chern-Simons theories  which includes,
for a particular choice of  parameters,
the Chern-Simons formulation
 of three-dimensional
gravity, with  general values of the cosmological constant
\cite{AT,Witten1}.  We explain how, for a different choice of parameters,
the theory can be used to construct a model of interacting
particles which is invariant  under the de Sitter version of 
$\kappa$-Poincar\'e symmetry. 
More generally, we consider the de Sitter and anti-de Sitter versions
of $\kappa$-Poincar\'e symmetry in three dimensions \cite{BHDS,BBH},
and clarify which of these
is associated with a Chern-Simons model. Our analysis shows clearly  that, for the usual timelike deformations,   
neither $\kappa$-Poincar\'e symmetry nor its de Sitter 
 and anti-de Sitter versions are  compatible with
3d gravity.  This conclusion is in disagreement  with  claims  in  the literature, for example  in the paper \cite{FKGS},
that $\kappa$-Poincar\'e symmetry does arise in 3d quantum gravity.
We discuss  this claim  and our reasons for disagreeing with it at the end of  our  
conclusion.

\subsection{Background:  Poisson-Lie groups, $r$-matrices and Hopf algebras}
\label{symreview}

 Before we
can explain our approach in more detail we need to
review some basic aspects of Hopf algebras and their classical
analogues, Poisson-Lie groups. Both provide generalisations of the
 usual implementation of a symmetry
 via a group.  The following summary is geared
towards spacetime symmetries in relativistic physics. For a more
general discussion and further details we refer the reader to
\cite{CP,Majid, K-S}.

In usual special-relativistic physics, the generators of the
(undeformed) Poincar\'e Lie algebra  arise in two, physically quite
distinct ways. In the first instance  they play the role of
infinitesimal symmetry generators of Minkowski space. As such they
exponentiate to elements of the symmetry group of Minkowski space
i.e.  the Poincar\'e group itself. However, we also come across the
Poincar\'e Lie algebra when we study the phase of a free
relativistic particle. Here, the Poisson brackets of the components
of momentum and  generalised angular momentum\footnote{We use
``generalised angular momentum'' to   mean both the angular momentum
components and the quantities which are conserved due to boost
invariance} reproduce the Lie algebra of the
Poincar\'e group. This is a familiar fact, which follows from
Noether's theorem. However, in the current context it is important
to keep in mind the different guises in which the Poincar\'e
generators appear
 - as generators of a symmetry group and
as coordinate functions on a particle phase space - and to
understand how they are related. It turns out that, mathematically,
the phase space of a free relativistic particle
is a coadjoint orbit in the dual of the Poincar\'e Lie
algebra. This makes it natural to think of
the Lie algebra generators as (linear) functions  on phase, and
shows that the Lie algebra of the symmetry and  the phase space are
dual to each other.

When considering classical systems corresponding to quantum
systems with quantum group symmetry, the
 two appearances of the symmetry  described
above become structurally richer while retaining their duality. The
symmetry group (generalising the Poincar\'e group above) gets
equipped with a Poisson structure, and the phase space (generalising
the particle phase space) becomes embedded in a group. Both the
symmetry group and the ambient group for the phase space become what
is known as a Poisson-Lie group - a space which is both a Lie
group and a Poisson manifold in such a way that these two structures
are compatible. Furthermore, the ``symmetry'' and ``phase space''
Poisson-Lie groups turn out to be dual to each other, in  a
mathematically precise way.

Poisson-Lie groups have associated infinitesimal structures, called
Lie bialgebras. They encode infinitesimal
versions of
 the Lie group structure via the  Lie bracket and of the
Poisson structure via an additional structure, called  co-commutator.
If two Poisson-Lie groups are in duality,
so are their Lie bialgebras: the commutator
of one determines the co-commutator of the other
and vice-versa. Thus, in the terminology of the previous paragraph, 
the commutator of the
``symmetry'' Poisson-Lie group agrees with
the  Poisson brackets  of the ``phase space''
Poisson-Lie group near the identity.

In most applications, the co-commutator of a Lie bialgebra is given
in terms of a special element of the tensor product of two copies of the Lie algebra, called the classical $r$-matrix. When this is the case,
the Lie bialgebra is called quasitriangular and  the Poisson
brackets of the associated Poisson-Lie group can be be expressed in
terms of the $r$-matrix. The resulting Poisson bracket is called the
Sklyanin bracket, and what was called ``symmetry'' Poisson-Lie group
is called Sklyanin Poisson-Lie group in this case. 
By duality, the $r$-matrix also fixes the commutator
of the dual Lie algebra, and hence the Lie
group structure of the dual or ``phase space'' Poisson-Lie group.
Thus, the knowledge of the original Lie brackets together with the
$r$-matrix is, in principle, sufficient to compute both the Sklyanin
Poisson-Lie structure and its dual. 

When the $r$-matrix 
satisfies   an additional non-degeneracy condition, one can 
 use it to define a
diffeomorphism between the original (``symmetry'') Poisson-Lie group
and its dual. Using this diffeomorphism, one can pull back the dual
Poisson structure to the original Poisson-Lie group, thus defining a
second Poisson structure on it. This Poisson structure is the
``phase space'' Poisson bracket, but written in terms of the
original Lie group. We shall refer to it as the dual Poisson
structure;  it can again be expressed in terms of the original
$r$-matrix.

The quantisation of classical systems with Poisson-Lie symmetry
typically leads to quantum systems whose symmetries are implemented by 
Hopf algebras \cite{CP}. In  a  precise sense, one can regard
the Hopf algebra as the quantisation of the Poisson-Lie group in that
case. Moreover, 
quantisation of mutually dual Poisson-Lie groups leads to
mutually dual Hopf algebras \cite{CP}.
Thus, for every Poisson-Lie structure on a given group  
one expects there
 to be two, mutually dual associated 
 Hopf algebras

In the case of the Poincar\'e group in four dimensions,  two dual Hopf
algebra deformations were, historically,  constructed directly, 
and not via quantisation of  a Poisson-Lie structure.
 The first version that was discovered \cite{LNRT} is now known as
the $\kappa$-Poincar\'e algebra;
the other, discovered shortly afterwards \cite{SWW,Majid2,Zakrzewski},
is often referred to as the $\kappa$-Poincar\'e  group. The duality
between the $\kappa$-Poincar\'e  group and $\kappa$-Poincar\'e algebra is
explicitly exhibited in  \cite{KM}.
 Both are sometimes  referred to as  $\kappa$-Poincar\'e
symmetries, and we followed that practice in our title and  abstract. 
In this paper we will also consider de Sitter and anti-de Sitter versions
of the $\kappa$-Poincar\'e algebra and group, and sometimes call them 
 the $\kappa$-(anti-) de Sitter  algebra and group. For detailed 
definitions and properties of these Hopf algebras in three dimensions
we refer the reader to  \cite{BHDS,BBH}.

Even though the $\kappa$-Poincar\'e algebra and the
$\kappa$-Poincar\'e group were not  constructed  via
quantisation, we can associate them to the classical
symmetries discussed above by taking the  classical 
limit. One then finds that the classical limit of the 
 $\kappa$-Poincar\'e algebra  is 
the ``phase space'' Poisson-Lie group described above, and that of 
the $\kappa$-Poincar\'e  group
 is the ``symmetry'' Poisson-Lie group. In practice, the 
calculation of the classical limit involves computing the first
order deformation in the co-product of $\kappa$-Poincar\'e algebra, and 
extracting a classical $r$-matrix. This is explained in detail in
\cite{BBH} for the $\kappa$-Poincar\'e algebra  and its
de Sitter and  anti-de Sitter versions in three dimensions.

\subsection{Chern-Simons theory and Poisson-Lie symmetries:
the Fock-Rosly construction}

\label{FRreview}

In general, it is difficult to construct physically motivated
and mathematically non-trivial phase spaces with
(non-trivial) Poisson-Lie  symmetries. In this paper we will use
a method developed by Fock and Rosly for such a construction.
Fock and Rosly showed in \cite{FR} that the phase space and
Poisson structure of Chern-Simons theory with gauge group $H$ on
manifolds of topology $\RR\times S$,
where $S$ is a closed, oriented two-surface, possibly
with handles  and punctures, can be
described in terms of an auxiliary Poisson structure.
If $S$ has $n$ punctures and $g$ handles, the auxiliary Poisson
structure is defined on the space $H^{n+2g}$;
the physical phase space of the associated Chern-Simons theory  with its canonical Poisson structure induced by the Chern-Simons action is obtained  after implementing a constraint and 
dividing by the conjugation action of 
$H$.

Fock and Rosly's auxiliary Poisson structure is defined in terms of a classical 
$r$-matrix, which is required to be compatible with the Chern-Simons
action in the following sense. 
Recall that a Chern-Simons action for the   gauge group $H$ 
requires for its definition an $\Ad$-invariant,
non-degenerate and symmetric bilinear form on the Lie algebra of $H$.
In this paper we call a  classical $r$-matrix
compatible with the Chern-Simons action if it lies in  the Lie algebra
of the group $H$ tensored with itself, satisfies the classical Yang-Baxter
equation and has a symmetric part which  
equals the Casimir associated to the  symmetric  
form used in the Chern-Simons action.
The $r$-matrix used in the Fock-Rosly construction
can be used to define  Lie bialgebras and Poisson-Lie groups,
as explained above. The compatibility requirement 
ensures in particular that the symmetric part of the $r$-matrix
is non-degenerate, so that the  (local) diffeomorphism
between the Sklyanin Poisson-Lie group $H$ and its dual
exists in this case.

The Fock-Rosly Poisson structure is such that
the Sklyanin Poisson-Lie group  $H$   is a symmetry i.e. 
acts via Poisson-isomorphisms on the auxiliary phase space.
 Moreover, it was
demonstrated by Alekseev and Malkin \cite{AMII} that the
contribution of different handles and punctures to this Poisson
structure can be decoupled and related  to standard Poisson
structures: each puncture
corresponds to a copy of the dual  Poisson structure on $H$ described above,
while each handle is related to a copy of the so-called
Heisenberg double Poisson structure \cite{CP,setishan}.
Thus, the Fock-Rosly Poisson-structure, and hence the Poisson structure on the physical phase space  of the associated Chern-Simons theory,
 is completely determined if the dual
and Heisenberg double Poisson structures are given.  

In physical applications, the surface $S$ is usually interpreted as 
``space'',
and the  punctures on it as particles. Thus, if the number of punctures is 
$n>1$,  the Fock-Rosly construction
leads to a Poisson algebra that  can serve as model for $n$ interacting
 particles.
The detailed physical interpretation of the phase space coordinates
can be quite involved; for details in the context of  the Chern-Simons formulation of 
3d gravity we refer the reader to \cite{BNR,we1,we5}.

\subsection{Overview of the paper}

After this brief review of Poisson-Lie theory, we can
summarise the paper  in more precise and technical terms. We begin,
in Sect.~2, with a review of the Lie algebras which arise as
infinitesimal local symmetries in 3d gravity with or without a
cosmological constant. We give a detailed discussion of the space of $\Ad$-invariant,
symmetric  bilinear forms on these Lie algebras   and derive a 
simple non-degeneracy criterion for such forms. We use a description
of the Lie algebras as the Lorentz (or, in the Euclidean case,
rotation) Lie algebra over a ring whose multiplication law depends on the cosmological
constant. This formalism, invented in \cite{ich3}, turns out to be
the most efficient way of treating all signs of the cosmological
constants and both the Lorentzian and Euclidean signature in a
unified way.

In Sect.~3 we discuss Chern-Simons theory with the local isometry
groups of 3d gravity as gauge groups, using the most general $\Ad$-invariant,
non-degenerate symmetric bilinear form of Sect.~2. The Chern-Simons actions
in this section were first considered  in \cite{Witten1} and  are  generalisations of the Chern-Simons formulation
of 3d gravity.  We
describe the coupling of the Chern-Simons field to point particles,
and show that the equations of motion in the absence
of particles are independent of the symmetric form, but that the
Poisson structure of phase space and the coupling to point particles
depend on it. We argue that, as a result, Chern-Simons theories for 
different choices of the non-degenerate symmetric bilinear form 
are physically inequivalent.
The observation about the variation of the Poisson
structure with the symmetric form was made, in different notation,
in \cite{BL}, where the similarity with the variation of the Immirzi
parameter in four-dimensional gravity was emphasised.

Sect.~4  deals with classical $r$-matrices, and 
 contains the main result of our paper.  We consider $r$-matrices
obtained from the $\kappa$-Poincar\'e algebra and its de Sitter and anti-de Sitter analogues \cite{BBH} via the classical limit   explained at the end of Sect.~\ref{symreview}; the $r$-matrices obtained in this way are all anti-symmetric.
 We then  check  if they  can be made  compatible
with the Chern-Simons action of Sect.~3, in the sense defined
in Sect.~\ref{FRreview}.   
This amounts to checking if the anti-symmetric
$r$-matrix obtained via the classical limit can be combined with 
 the (symmetric) Casimir corresponding to the $\Ad$-invariant, non-degenerate
symmetric  bilinear form
used in the Chern-Simons action to give a solution of 
the classical Yang-Baxter equation. We show that this is the 
case provided a vector appearing in the anti-symmetric part of the 
$r$-matrix satisfies a certain condition, which we discuss.
 One finding, worth stressing at this stage, is  
that the classical $r$-matrix obtained from the $\kappa$-Poincar\'e
algebra  is only compatible with   the Chern-Simons action of 3d gravity if
the 
vector appearing in the $r$-matrix is taken
to be spacelike. Since this vector is timelike
in the usual form of the  $\kappa$-Poincar\'e
algebra,  we conclude that the usual $\kappa$-Poincar\'e  symmetry 
is not associated with  3d gravity.

In Sects.~5 and 6 we compute  Lie bialgebra and Poisson-Lie
structures associated to the $r$-matrices of Sect.~4. We compute and
discuss the Sklyanin and dual Poisson brackets, focussing on the de Sitter
case and the situation where the  special vector appearing in the
$r$-matrix is timelike. We obtain  general formulas, valid for any
value of the cosmological constant, and discuss both their linearisation
near the identity, and their limit as $\Lambda\rightarrow 0$. 

Sect.~7 contains our conclusions and
outlook. We explain how our calculations enable one 
 to construct multi-particle models which are invariant under the Poisson-Lie
group which arises as the classical limit  of the $\kappa$-de Sitter group. We end with comments on 
  the relation  between $\kappa$-Poincar\'e symmetry and gravity in 
three dimensions.

\section{Local isometry groups and their Lie algebras}

In discussions of three-dimensional spacetimes 
we adopt the  notational conventions of \cite{we6}, which we briefly 
review. We set the speed of light to $1$, and  write
$\eta^E=\text{diag}(1,1,1)$  for the three-dimensional Euclidean metric
and  $\eta^L=\text{diag}(1,-1,-1)$ for the three-dimensional Minkowski
metric; we omit the superscript in formulas valid for both signatures. In
particular, we use the abbreviations
\begin{align}
\bp\cd\bq=\eta_{ab}p^aq^b,\qquad\mbox{with}\quad \bp=(p^0,p^1,p^2),\;\;
\bq=(q^0,q^1,q^2)\in\RR^3,
\end{align}
 as well as  $\bp\bq$ for $\bp\cd\bq$
 and $\bp^2$ for $\bp\cd \bp$.

We denote by $J_a$, $a=0,1,2$,
 the generators of both the three-dimensional
rotation algebra $\mathfrak{su}(2)$ and the three-dimensional
Lorentz algebra $\mathfrak{su}(1,1)$, and use the letter $\gothh$
 for either of these Lie algebras. The Lie brackets are
\begin{align}
\label{lorbr} [J_a,J_b]=\epsilon_{abc}J^c,
\end{align}
where $\epsilon$ denotes the fully
antisymmetric tensor in three dimensions with the convention
$\epsilon_{012}=\epsilon^{012}=1$ (for both signatures),
and the indices are raised with $\eta^E$ in the Euclidean case,
and with $\eta^L$ in the Lorentzian case. In the Lorentzian case,
 $J_0$ is the generator of the spatial rotations and $J_1,J_2$
are the generators of the boosts.
As explained in
the appendix, the matrix of the 
 Killing form on $\gothh$
 turns out to be  $-2\eta_{ab}$ in this basis; dividing the Killing form by the factor $-2$ we obtain the 
invariant, non-degenerate bilinear form $\eta$ satisfying
 \bea
\label{etadef}
\eta (J_a,J_b)=\eta_{ab}.
\eea

In 3d gravity, solutions of the Einstein solution are locally
isometric to certain ``model spacetimes'' which are completely
determined by the signature of spacetime and the cosmological
constant. The isometry groups of these model spacetimes are
hence local isometries of 3d gravity, a situation which differs
considerably from the four-dimensional case where one has only local
Lorentz symmetry.
We list the groups we want
to consider in Table 1. If we do not need to specify the group, we
use $H$ to stand for any of the groups in the table.

\vbox{
\baselineskip 38 pt
\begin{center}
\begin{tabular}{|c|c|c|}
\hline
$\Lambda_c$   & Euclidean signature & Lorentzian signature \\
\hline
$ =0$  & $SU(2)\ltimes \RR^3$ & $SU(1,1)\ltimes \RR^3$ \\
\hline
$> 0$ &
$ SU(2)\times SU(2)$&
$SL(2,\CC)$ \\
\hline
$ < 0$ & $ SL(2,\CC) $&
$SU(1,1)\times SU(1,1) $\\
\hline
\end{tabular}
\vspace{0.1cm}

Table 1: Local isometry groups in 3d gravity
\end{center}
}

\baselineskip  16 pt

The local  isometry groups  have Lie algebras which
 can be expressed in a unified fashion, with
the cosmological constant $\Lambda_c$  playing the role of a deformation
parameter \cite{Witten1}. Defining
\bea \label{Lambdadef} \Lambda = \left\{
\begin{array}{l l}
 \Lambda_c  & \mbox{for Euclidean signature} \\
   -\Lambda_c &\mbox{for Lorentzian signature}
\end{array}
\right.
\eea
 these Lie algebras, in the following denoted by
$\gothh_\Lambda$, are the six-dimensional Lie algebras with
generators $J_a,P_a$, $a=0,1,2$, and Lie brackets\footnote{Our parameter $\Lambda$ is called $\lambda$ in
\cite{Witten1}.}
\begin{align}
\label{liebra} [J_a,J_b]=\epsilon_{abc} J^c,\qquad
[J_a,P_b]=\epsilon_{abc} P^c,\qquad
[P_a,P_b]=\Lambda\epsilon_{abc}J^c.
\end{align}
For  $\Lambda=0$, the bracket of the generators $P_a$
vanishes, and the Lie algebra $\gothh_\Lambda$ is the
three-dimensional Euclidean and Poincar\'e algebra. For $\Lambda<0$,
the brackets  \eqref{liebra} are those of the Lie algebra
$\mathfrak{sl}(2,\CC)$ for both  Euclidean and Lorentzian signature.
For $\Lambda>0$ the brackets are those  of $\mathfrak{su}(2)\oplus \mathfrak{su}(2)$ in the Euclidean and of
$\mathfrak{su}(1,1)\oplus \mathfrak{su}(1,1)$ in the Lorentzian case.


For a unified description of  the local isometry groups and their
Lie algebras it is convenient to use a trick which was discovered in
\cite{ich3} and used extensively in \cite{we6}. The idea is to
introduce a formal parameter $\tl$ which satisfies $
\tl^2=\Lambda$   and to identify the generators $P_a$ in
\eqref{liebra} with $\theta J_a$. It is easy to check that the
brackets \eqref{liebra} then follow from \eqref{lorbr} by extending
\eqref{lorbr} linearly over $\theta$.

As explained in  \cite{ich3} this construction amounts to considering the
 commutative ring $R_\Lambda$ consisting  of elements of the form
$a+\theta b$, $a,b\in \RR$ and viewing the Lie algebra
$\gothh_\Lambda$ as  the realification of the Lie algebra $\gothh$ tensored with  $R_\Lambda$.
We refer to \cite{ich3} for a formal definition and details, but recall the notation
\begin{align}
\label{imredef} \text{Re}_\tl(a+\tl b)=a\qquad \text{Im}_\tl(a+\tl
b)=b\qquad \forall a,b\in\RR,
\end{align}
and the conjugation
\begin{align}
\label{ringconj} \overline{(a+\tl b)}=a-\tl b.
\end{align}
The ring $R_\Lambda$
is a field in the case $\Lambda <0$ (the complex numbers)
but  has zero divisors when $\Lambda \geq 0$, as we shall see further below.

To illustrate the use of the parameter $\theta$, consider the
$\Ad$-invariant symmetric bilinear forms on $\gothh_\Lambda$. As
pointed out in \cite{Witten1} there
is a two-dimensional vector space of such forms,  with  a basis is given by
\begin{align}
\label{pair} &t( J_a,J_b)=0, & &t(P_a,P_b)=0, &
&t( J_a,P_b)=\eta_{ab},\\
\label{othform} &s(J_a,J_b)=\eta_{ab}, & &s(J_a,P_b)=0, &
&s(P_a,P_b)=\Lambda\eta_{ab}.
\end{align}
It was shown in \cite{ich3} that these forms
can be obtained as the real and imaginary part
of the invariant, symmetric bilinear form $\eta$ \eqref{etadef} on $\gothh \otimes R_\Lambda$: for $X,Y\in \gothh_\Lambda$ we have
\bea
\label{stdef} s(X,Y)=\text{Re}_\tl(\eta(X,Y))\qquad
t(X,Y)=\text{Im}_\tl(\eta(X,Y)),
\eea 
where $P_a$ should be interpreted as $\theta J_a$ when it appears on the 
right-hand side.
 More generally, we can consider linear combinations of
the forms $s$ and $t$, which we write as 
\bea
(\cdot, \cdot)_\tau=  \al t(\cdot,\cdot)+ \bet s(\cdot,\cdot)  =
\text{Im}_\tl(\tau\eta(\cdot,\cdot)), \eea 
with $\tau = \al +\theta \bet$.
Explicitly
\bea
(J_a, J_b)_\tau=\beta \eta_{ab}, \quad (J_a, P_b)_\tau=\alpha \eta_{ab}, \quad (P_a, P_b)_\tau=\Lambda \beta \eta_{ab}. 
\eea
For the construction of  Chern-Simons actions in the next section
we require an $\Ad$-invariant,
symmetric bilinear form on $\gothh_\Lambda$ which is non-degenerate.
The following lemma gives a simple criterion for the  non-degeneracy of
$( \cdot ,\cdot )_\tau$.

\begin{lemma}
 The $\Ad$-invariant, symmetric 
bilinear form $(\cdot,\cdot)_\tau$ is non-degenerate iff 
\bea 
\label{taucond}
\tau
\bar\tau= \al^2-\Lambda \bet^2 \neq 0. 
\eea 
\end{lemma}

Note that the condition \eqref{taucond} is
always satisfied (for non-zero $\tau$) if $\Lambda <0$ but that it is
non-trivial in the other cases.

{\bf Proof}: \quad Recall that a non-degenerate bilinear
form on a vector space establishes an isomorphism between the vector
space and its dual. We will need this map later, so  we show
that $(\cdot,\cdot)_\tau$ is non-degenerate if \eqref{taucond} holds
by explicitly giving the map  $\isom:\gothh_\Lambda^*\mapsto \gothh_\Lambda$ which 
satisfies
\bea
\label{assdef}
\xi(X)=(\isom(\xi), X)_\tau, \quad \forall X\in \gothh_\Lambda.
\eea
and by  showing that it
is bijective.
Consider the basis
\bea
\label{basis}
B=\{J_0,J_1,J_2,P_0,P_1,P_2\}
\eea
of $\gothh_\Lambda$  and the
dual basis of  $\gothh_\Lambda^*$:
\bea
\label{dualbasis}
B^* =\{J^*_0,J^*_1,J^*_2,P^*_0,P^*_1,P^*_2\}.
\eea
It is easy to check that \eqref{assdef}  is satisfied if we set
\bea
\label{Omdef}
\isom(J^*_a)&=&\frac{\theta}{\tau}J_a= \frac{1}{\al^2-\Lambda \bet^2}(\alpha P_a-\Lambda \beta J_a)\nonumber \\
\isom(P^*_a)&=&\frac{1}{\tau}J_a= \frac{1}{\al^2-\Lambda \bet^2}(\alpha J_a- \beta P_a),
\eea
showing that $\isom$ is well-defined if and only if $\al^2-\Lambda \bet^2=\tau \bar\tau\neq 0$. The inverse is given by
\bea
\isom^{-1}(J_a)=(\alpha P^*_a+ \beta J^*_a)\qquad
\isom^{-1}(P_a)=(\alpha J^*_a+ \beta \Lambda P^*_a).
\eea
Thus $\isom$ exists and is invertible iff $\tau\bar\tau\neq 0$,
as claimed.
\hfill $\Box$

\section{Chern-Simons action,
Poisson structure and the coupling to  particles}
\label{actsect}

A  Chern-Simons theory on a three-dimensional manifold depends
for its definition on a choice of gauge group and an $\Ad$-invariant, non-degenerate
symmetric bilinear form
on the Lie algebra of that gauge group. Remarkably, as was shown in  \cite{AT,Witten1},  
one obtains
the Einstein-Hilbert action for three-dimensional gravity for
any signature and value of the cosmological constant from the
Chern-Simons action by picking the appropriate local isometry group
from Table~1 as gauge group and using the non-degenerate form $t(\cdot,\cdot)$
\eqref{pair}. In this section we consider what happens if, instead
of $t(\cdot,\cdot)$, we use the more general non-degenerate form $(\cdot,\cdot)_\tau$
to define a  Chern-Simons theory.
We couple the gauge field to  matter in the form of point particles, and study the resulting phase space and its Poisson structure. 
We assume basic facts about Chern-Simons theory, and refer the 
reader to \cite{we4} for a more detailed treatment in a related context.
The generalised Chern-Simons action that we study here was already considered 
by Witten in \cite{Witten1}, and 
some of the results regarding Poisson brackets  were derived in different notation
in \cite{BL}. However, the coupling to point particles does not appear to have 
been considered elsewhere.  In the context of this paper it is important to understand
to what extent 
the variation of the form $(\cdot,\cdot)_\tau$ leads to physically
inequivalent theories. We comment on this issue at the end of this section.

We work on
 a  three-manifold $M$ (``spacetime'') of topology
 $\RR\times S$, where $S$ is an oriented two-dimensional
manifold  of genus $g$ with possible punctures. The punctures
are needed in order to introduce matter in the form of
point particles into the model.
We also need a coordinate $x^0$ on $\RR$,
and write  $x=(x^1,x^2)$ for local coordinates on $S$.
To keep our formulas simple we  consider only one puncture with coordinate
$x_*$ on $S$;
the generalisation to several punctures is straightforward \cite{we4}.
 The  gauge field of the Chern-Simons theory     is locally
a one-form on spacetime with values in the Lie algebra
$\gothh_\Lambda$ of one of the isometry groups in Table~1.
In terms of the generators $J_a$ and $P_a$ we have the expansion
\bea
\label{Adec}
A=\omega_aJ^a + e_aP^a,
\eea
where $
\omega= \omega^aJ_a$
is geometrically interpreted as  the spin connection on the
frame bundle and
the set of one-forms
$\{e_0,e_1,e_2\}$  as a dreibein
(provided it is invertible).
The curvature  of $A$  is
\bea
\label{Fdec}
F=dA+\frac 1 2 [A\wedge A]=R+C+T,
\eea
and contains the Riemann curvature
\bea
R= d\omega + \frac 1 2 [\omega\wedge\omega],
\eea
the cosmological term
\bea
C=\frac \Lambda 2\epsilon^{abc} e_a\wedge e_b J_c,
\eea
and the  torsion
\bea
\label{torsion}
T=(de^c +\epsilon^{abc}\omega_a\wedge e_b) P_c.
\eea

Using the  product structure  $M=\RR\times S$ we
decompose the gauge field as
\begin{align}
\label{gauge field}
A=A_0 dx^0+A_S,
\end{align}
where $A_S$ is an $x^0$-dependent and Lie algebra valued
one-form on $ S$ and $A_0$ is a Lie algebra valued function on
$\RR\times S$.
We use the usual notation
$d$ for  the  exterior derivative  on $\RR\times S$ and write
$d_S$ for the exterior derivative on $S$.
With this notation, the field strength two-form can be decomposed as
\bea
\label{curvdec}
F=dA + A\wedge A = dx^0\wedge(\partial_0 A_S - d_S A_0 + [A_0,A_S]) + F_S,
\eea
where $F_S$ is the  curvature two-form on $S$:
\bea
\label{curvv}
F_S=d_SA_S + A_S\wedge A_S.
\eea
The   Chern-Simons action for the gauge field $A$ is
\bea
\label{acti}
I_\tau(A)=\int_M (A\wedge dA)_\tau + \frac 1 3 (A\wedge[A,A])_\tau.
\eea
In order to read off the  constraints and the symplectic structure defined by this action, 
it is useful to perform a (2+1)-decomposition of the action:
\bea
\label{actio}
 I_\tau[A_S,A_0]=
 \int_\RR dx^0\int_{S}(\partial_0 A_S\wedge A_S)_\tau +
( A_0\,,\, F_S )_\tau.
\eea
Variation with respect to $A_0$, which acts as a Lagrange multiplier, gives the constraint
\bea
\label{aconst}
 F_S(x)=0,
\eea
and variation with respect to $A_S$ gives the evolution equation
\bea
\label{aeom}
\partial_0 A_S=d_SA_0+[A_S,A_0].
\eea
Together, these two equations are equivalent to the statement that the
field strength $F$ is zero. Geometrically this means that
the torsion $T$ vanishes, and  that  the curvature is constant:
\bea
T=0,\qquad\quad R+C=0.
\eea
 Note, in particular, that
both the constraint and the evolution  equation 
  are independent of the choice of $(\cdot,\cdot)_\tau$.
One might think that this means that theories corresponding to 
different choices of $(\cdot,\cdot)_\tau$ are physically equivalent.
However, this is not case, 
as we shall explain at the end of this section.

To see how the theory is affected by changing  $\tau$, and
to gain a better understanding of the physical interpretation of the
action \eqref{acti} we  use the decomposition \eqref{Adec}. 
After integration by parts 
and dropping a boundary term the 
action \eqref{acti}  can  be written as  
\begin{align}
\label{action}
I_\tau(A)&=\alpha \int_M\left(2  e^a\wedge R_a  + \frac \Lambda 3 \epsilon_{abc}
e^a\wedge e^b\wedge e^c \right)\nonumber \\
&+\beta \int_M\left(\omega^a\wedge d\omega_a +\frac{1}{3}\epsilon_{abc}\omega^a
\wedge \omega^b\wedge\omega^c + \Lambda e^a\wedge T_a\right),
\end{align}
where $T_a$ are the components of the torsion \eqref{torsion}.
Note that  first line contains the usual action for 3d gravity with cosmological
constant $\Lambda$. The first term in the second line is simply the Chern-Simons action 
for the spin connection $\omega$, with $\omega$ treated as an
independent variable. The formula \eqref{action}
also shows clearly that in the non-gravitational case $\alpha=0$ the action
becomes  independent of the dreibein $e_a$ and 
hence degenerate in the limit $\Lambda \rightarrow 0$.

In order to study the symplectic structure associated with this action,
we  write out the terms in  \eqref{actio}:
\begin{align}
\label{actionn}
 I_\tau[A_S,A_0]&=
 \alpha
 \int_\RR dx^0\int_{S} \eta_{ab}(\partial_0 e^a_S\wedge \omega^b_S
+\partial_0\omega_S^a\wedge e_S^b) \nonumber \\
&+\beta \int_\RR dx^0\int_{S} \eta_{ab}(\partial_0 \omega^a_S\wedge \omega^b_S
+\Lambda \partial_0e_S^a\wedge e_S^b)\nonumber \\
&+\qquad \text{constraint}.
\end{align}

This expression is instructive in a number of ways. First of all,
it allows us to determine the physical dimensions of the coupling
constants $\alpha$
and $\beta$.  The dimension
of the dreibein $e_a$ is length and the spin connection $\omega$
is dimensionless; thus, working in units where the speed of light is 1, the constant $\alpha$ has to have the dimension of mass in order for the action to
have the correct dimension. In the usual gravitational interpretation
of the action \eqref{action} with $\beta=0$, $\alpha$ is identified with the
inverse of the gravitational constant $G$ in three dimensions. Recalling that
the cosmological constant has the dimension of inverse length squared, we
see that the terms in the second row of \eqref{actionn} have the correct
dimension if $\beta$ has the dimension of an action i.e. mass times length.

The second important lesson we learn from
\eqref{actionn} is that the symplectic structure of the theory
depends on $\tau$. Specifically, we can read off from \eqref{actionn}
 which fields are canonically conjugate to each other for given $\tau$.
We refer the reader to \cite{BL}
where  the general formulas for
 conjugate variables are given (in different notation) and
the similarity between the parameters $\alpha$ and $\beta$ 
and the Immirzi parameter
in four-dimensional gravity is pointed out.
For our purposes it is worth emphasising two cases. For the gravitational
form $t$  (i.e. $\tau  = \alpha$) the spatial components of the dreibein $e^a$
are conjugate to the spatial components
of the spin connection. With $e^a_S=e^a_idx^i,$  and $\omega^a_S=\omega^a_idx^i$ the only non-vanishing brackets are 
\bea
\{e_i^a(x),\omega^b_j(y) \}=\frac{1}{2\alpha}
\eta^{ab}\epsilon_{ij}\delta^{(2)}(x-y).
\eea
 In the extreme non-gravitational case $\tau =\theta\beta$,
the dreibein and the spin connection are self-conjugate:
\bea
\label{eebr}
\{e_i^a(x),e^b_j(y) \}=\frac{1}{2\Lambda\beta}
\eta^{ab}\epsilon_{ij}\delta^{(2)}(x-y), \quad
\{\omega_i^a(x),\omega^b_j(y) \}=\frac{1}{2\beta}
\eta^{ab}\epsilon_{ij}\delta^{(2)}(x-y).
\eea

The way geometry is coupled to matter in Einstein gravity can be
emulated in the gauge theory formulation by the following procedure:
each puncture is decorated with the action of a free (relativistic)
particle moving in the model spacetime, and this action is minimally
coupled to the gauge field. The action of a free particle in the
model spacetime has, in turn, a simple description in terms of
co-adjoint orbits of the local isometry group $H$. In the current
context it is worth stressing that no invariant, non-degenerate symmetric form
is needed on the Lie algebra $\gothh_\Lambda$ of $H$ in order to
define the free particle action. However, in order to derive the
equations of motion of the combined gauge theory - particle system,
we do require such a form. The form thus describes the coupling
of the particle and the bulk degrees of freedom. We therefore
formulate the particle action in terms of the form
$(\cdot,\cdot)_\tau$ from the outset.

For our discussion we consider one particle, with mass
 $m$ and spin $s$.  We encode both in one
element $\hat \xi$ of the dual Lie algebra $\gothh_\Lambda^*$
via \bea \hat \xi=m P^*_0 + sJ^*_0. \eea The momentum $p_a$ and
generalised angular  momentum $j_a$  
of the particle in a general state of motion is
obtained via the co-adjoint action of an element   $h\in H$ on  $\hat
\xi$: \bea \xi=\Ad^*(h)(\hat \xi)=p^a P^*_a+j^a J^*_a; \eea 
their Poisson
brackets reproduce the  Lie brackets \eqref{liebra}:
\bea
\{j_a,j_b\}=\epsilon_{abc}j^c, \quad\{j_a,p_b\}=\epsilon_{abc}p^c,\quad
\{p_a,p_b\}=\Lambda\epsilon_{abc}j^c.
\eea

The Lie algebra elements $\isom(\hat \xi)$ and $\isom(\xi)$
associated to $\hat\xi$ and $\xi$ via \eqref{assdef} are related
by the adjoint action:
\bea
\label{tidef} \isom(\xi)=h \isom(\hat\xi) h^\inv.
\eea
 The combined field and particle action now takes the following form:
\begin{align}
 I_\tau[A_S,A_0, h]=
\label{ordaction}
 &\int_\RR dx^0\int_{S}(\partial_0 A_S\wedge A_S)_\tau
 -\int_\RR dx^0 (\isom( \hat\xi)\,,\, h^\inv\partial_0
h)_\tau\\
 +&\int_\RR dx^0\int_{S}( A_0\,,\, F_S -
 \isom(\xi)\delta^{(2)}(x-x_*)dx^1\wedge dx^2 )_\tau.
\nonumber
\end{align}
 Varying with respect to the Lagrange multiplier $A_0$
 we obtain the constraint
\begin{align}
\label{vara0equ} F_S(x)= \isom(\xi)\delta^{(2)}(x-x_*)dx^1\wedge dx^2.
\end{align}
Variation with respect to $A_S$ gives again the evolution equation
\eqref{aeom}. Together with \eqref{vara0equ} this means that the 
curvature $F$ vanishes except at the ``worldline'' of the puncture,
where it is given by the constraint \eqref{vara0equ}.
In order to interpret the constraint
  geometrically and physically 
we  use the decomposition \eqref{Fdec} of the curvature
and the map \eqref{Omdef}:
\begin{align}
R_S^a+C_S^a&=\frac{1}{\tau\bar\tau}\left(\alpha p^a -\Lambda \beta j^a\right)
\delta^{(2)}(x-x_*)dx^1\wedge dx^2\nonumber \\ T_S^a&=\frac{1}{\tau\bar\tau}\left(\alpha j^a -\beta p^a\right)
\delta^{(2)}(x-x_*)dx^1\wedge dx^2
\end{align}
The extreme cases are again most easily interpreted: in the gravitational
case $\tau=\alpha$, the momentum is a source of curvature and the generalised angular momentum is a source of torsion.
In the non-gravitational case $\tau =\theta \beta$, the momentum is a source of torsion and the generalised 
angular momentum a source of curvature. In the generic case, both momentum and generalised 
angular momentum are sources of both torsion and curvature.

To end this section we discuss  how the physics   of the  Chern-Simons theory  with action \eqref{ordaction}
  depends on the parameter $\tau$. We begin with an observation
regarding the equations of motion in the absence of matter. As we saw (and was previously stressed by Witten in \cite{Witten1})
 the  classical equation of motion for the gauge field, 
combining our constraint \eqref{aconst} and evolution equation  \eqref{aeom}, is  the flatness condition $F=0$ {\em regardless}
of which non-degenerate form  $(\cdot, \cdot)_\tau$ is used in the action. However, we shall now argue that 
the equation of motion
does not capture all of the physics, even classically. The constraint \eqref{aconst} states that, for any value of $x^0$, 
the restriction of  the connection $A$ to the surface $S$ is flat, while the  
evolution equation \eqref{aeom} simply states that the $x^0$-evolution
is via gauge transformations. Together, they imply that the physical phase space, as a manifold,  is the space of flat $H$-connections on $S$. Thus, it is indeed true that the phase space   manifold does not depend on 
the non-degenerate form $(\cdot, \cdot)_\tau$. 

However, any physical interpretation of the phase space and of functions on it (classical observables)  also depends on the 
 symplectic structure of the phase space.  This is inherited from the Poisson brackets of the gauge field which,
as we explained in the discussion preceding \eqref{eebr}, {\em does} depend on $(\cdot, \cdot)_\tau$.
In the current paper
 we can only illustrate  how the symplectic structure enters the interpretation of  the space of flat $H$ connections in terms
of spacetime geometries, and refer to  \cite{ich3,CM1,CM2} for details. As explained there in the context of  
 the usual gravitational action ($\beta=0$) for general values of the constant  $\Lambda$, one can associate to every closed  geodesic on the spatial surface  two canonical phase space  functions computed
from the   holonomy along the geodesic.  These functions generalise the masses and spins associated to point particles and generate geometrical transformations via Poisson brackets on the phase space. For $\beta=0$ the  ``mass variable" generates grafting (cutting a spatial surface along the geodesic and inserting a cylinder)
and  the ``spin variable'' generates earthquakes (cutting a spatial surface along the geodesic and rotating the edges of the cut
against each other).  These transformations can be viewed, respectively, as a translation and a rotation associated to a geodesic which supports the interpretation of their Hamiltonians as, respectively,  a momentum or mass and angular momentum or spin variable associated to the geodesic. 
If one  studied the Chern-Simons theory with the same gauge group but  with $\alpha=0$
and $\beta\neq 0$, the  physical role of ``mass'' and ``spin'' would be reversed \cite{CM1}, so the same phase space 
function now has a different physical interpretation.

The dependence of the physics on $(\cdot, \cdot)_\tau$ becomes even clearer  when 
one considers a  manifold with boundary or when one couples the gauge field to matter. In the Chern-Simons formulation
of 3d gravity, a boundary at ``spatial infinity'' can be modelled as a non-standard puncture \cite{we5}. There are  phase
space functions
 associated to that puncture  which are  interpreted as the ``total mass of the universe''  and ``total spin of the universe''. Via the Poisson brackets on the phase space induced by  $(\cdot, \cdot)_\tau$, with $\beta=0$, these phase space functions generate, respectively,  time evolution and rotation 
 of the universe  relative to a chosen centre-of-mass frame.
If, instead,
one sets $\alpha=0$, keeping $\beta\neq 0$, the roles of those functions are reversed, with the total spin
now generating time translations, and the total mass rotations.
This effect is the analogue of what we observed above for the coupling of the gauge field to point particles.
In the theory with $\alpha\neq 0, \beta=0$ the particle's momentum is a source of curvature and its generalised angular momentum is a source
of torsion, as expected  in any formulation of gravity. However, in the Chern-Simons theory with $\alpha =0,\beta \neq 0$
the roles of momentum and generalised angular momentum are reversed, leading to a coupling between matter and geometry 
which is quite different from what happens in  gravity. 

All these considerations show that Chern-Simon theories with different values of $\tau$ are physically
inequivalent. Even though the phase space is independent of $\tau$ {\em as a manifold}, the symplectic structure
and hence the physical interpretation depends on $\tau$. Generally speaking, we can summarise the above discussion
by saying that exchanging the gravitational case  $\alpha\neq 0, \beta =0$ with the non-gravitational case
 $\beta\neq 0, \alpha =0$ amounts to exchanging the roles of 
momentum and generalised angular momentum. Note, finally, that  one can also detect the difference between
these two cases by taking the limit $\Lambda \rightarrow 0$: in the gravitational case this limit exists,
in the non-gravitational case it does not.

\section{Classical r-matrices and their compatibility with Chern-Simons gauge theory}
\label{rsect}

As explained in the introduction, classical $r$-matrices provide
the link between Chern-Simons theory on the one hand and Hopf
algebras on the other. An  $r$-matrix  is associated 
to a given Chern-Simons theory via the Fock-Rosly construction,
which we reviewed in Sect.~\ref{FRreview}. Recall that, in that review,
we called an $r$-matrix compatible with a Chern-Simons action if it 
satisfies the classical Yang-Baxter equation and 
its symmetric part is equal to the Casimir associated to the $\Ad$-invariant,
non-degenerate symmetric bilinear form  used in the Chern-Simons action.
The connection between $r$-matrices and Hopf algebras is established
by taking a classical limit, as explained in Sect.~\ref{symreview}.
In this section we will establish a condition for the $r$-matrix obtained via
the classical limit of the $\kappa$-Poincare algebra and its de Sitter and anti-de Sitter version to be compatible with Chern-Simons action \eqref{action}.

We begin by fixing some terminology and notation.
The  classical Yang-Baxter equation for a Lie algebra $\gothg$ is an equation 
for an element $r=r_{mn}X^m\otimes X^n 
\in\gothg\otimes\gothg$.  With the standard  notation 
$r_{12}=r_{mn}X^m\otimes X^n \otimes 1$,$r_{13}=  r_{mn}X^m\otimes 1\otimes  X^n 
$, $r_{23}=  r_{mn}1\otimes X^m\otimes  X^n 
$ the equation reads \cite{CP} 
\bea
[[r,r]]:=[r_{12},r_{13}]+[r_{12},r_{23}]+[r_{13},r_{23}]=0. 
\eea
If the right-hand side is not zero but  an invariant element of 
$\gothg\otimes\gothg\otimes \gothg$ we say that $r$ satisfies
the modified classical Yang-Baxter equation.

Now  consider the 
anti-symmetric elements $r_A\in \gothh_\Lambda\otimes \gothh_\Lambda$
 which are obtained in \cite{BBH} from 
 the $\kappa$-Poincar\'e algebra and its de Sitter
and anti-de Sitter analogues in three dimensions by taking the 
classical limit, or, more precisely, by looking at the first-order
terms in the co-product of those Hopf algebras.
The formulas given  in \cite{BBH}  contain implicitly 
the unit timelike vector $(1,0,0)$; we generalise them  by considering 
instead an arbitrary vector $\bm$. In terms of the element
$M=m_aJ^a\in \gothh$  our generalisation of the $r$-matrices in \cite{BBH}
can be written as  
\bea
\label{rmat1}
r_A=J_a\otimes [M,P^a] + P_a\otimes [M,J^a].
\eea
We have the following
\begin{lemma}
For any $M=m_aJ^a\in \gothh$, the antisymmetric element
$r_A$ in \eqref{rmat1}
satisfies
\bea
\label{rasym}
[[r_A,r_A]]=\bm^2\epsilon_{abc}(\Lambda J^a\otimes J^b\otimes J^c+J^a\otimes P^b\otimes P^c
+P^a\otimes J^b\otimes P^c+ P^a\otimes P^b\otimes J^c)
\eea
\end{lemma}

{\bf Proof}: \quad
This can be shown by a lengthy but direct computation.
Alternatively, it can be deduced
 from the equation \eqref{use4} derived in the appendix  by
multiplying both sides with
$$\Lambda 1\otimes 1\otimes 1 + 1\otimes \theta\otimes\theta +
\theta\otimes 1\otimes \theta + \theta\otimes \theta \otimes 1.$$
\hfill $\Box$

Note that  the right-hand side
of \eqref{rasym} is an invariant element of
 $\gothh_\Lambda\otimes \gothh_\Lambda\otimes\gothh_\Lambda$; hence
 $r_A$ satisfies the
modified classical Yang-Baxter equation.
This is expected since
one can  show quite generally that the first-order terms
from which $r_A$ was obtained   define a co-commutator on $\gothh_\Lambda$ 
which gives it the structure of  
a Lie bialgebra.
 The modified classical Yang-Baxter equation
is precisely the condition for   $r_A$  to give rise 
to a Lie bi-algebra structure on $\gothh_\Lambda$.

Next we turn to  
the quadratic Casimir element  associated
to the metric $(\cdot,\cdot)_\tau$ used in defining the Chern-Simons action.
The general form of that Casimir element is
\bea
K_\tau = \frac{\alpha }{\tau\bar \tau}(J_a\otimes P^a+P_a\otimes J^a)
-\frac{\beta}{\tau\bar\tau}(\Lambda J_a\otimes J^a + P_a\otimes P^a).
\eea
In particular, for $\tau =1$ we obtain the Casimir associated to
the form $s$
\bea
K_s=J_a\otimes P^a+P_a\otimes J^a,
\eea
and for $\tau =\theta$ we obtain the Casimir associated to
the form $t$
\bea
K_t=J_a\otimes J^a+\frac{1}{\Lambda}P_a\otimes P^a.
\eea

In order to check the compatibility of $r_A$ with the Chern-Simons 
action we need to check  if 
\bea
\label{rmatt}
r=K_\tau +r_A
\eea
satisfies the classical Yang-Baxter equation. As explained
on p 54 of \cite{CP},
it follows
from the $\gothh_\Lambda$-invariance of $K_\tau$ that for any
anti-symmetric element $r_A\in \gothh_\Lambda\otimes\gothh_\Lambda$
\bea
\label{cpneat}
[[K_\tau+r_A,K_\tau+r_A]]=
[[K_\tau,K_\tau]]+[[r_A,r_A]].
\eea
Thus we need to check if 
$[[r_A,r_A]]=-[[K_\tau,K_\tau]]$.
A straightforward  calculation shows
\begin{align}
\label{rsym}
&[[K_\tau,K_\tau]]=\frac{\alpha^2+\Lambda \beta^2}{(\tau\bar\tau)^2}
\epsilon_{abc}(\Lambda J^a\otimes J^b\otimes J^c+J^a\otimes P^b\otimes P^c
+P^a\otimes J^b\otimes P^c+ P^a\otimes P^b\otimes J^c)\nonumber \\
&\quad-\frac{\alpha\beta}{(\tau\bar\tau)^2}
\epsilon_{abc}( P^a\otimes P^b\otimes P^c+\Lambda P^a\otimes J^b\otimes J^c
+\Lambda J^a\otimes J^b\otimes P^c+\Lambda  J^a\otimes P^b\otimes J^c).
\end{align}

Combining \eqref{rsym} and \eqref{rasym} we arrive at
\begin{theorem}
\label{rresult}
The element $r=K_\tau+r_A\in \gothh_\Lambda\otimes \gothh_\Lambda$ satisfies the
classical Yang-Baxter equation iff
\bea
\label{match}
\bm^2=-\frac{1}{\tau^2}
\eea
or, writing real and imaginary components explicitly,
\bea
\label{Ncond}
\alpha\beta =0\qquad \mbox{and}\qquad \bm^2= -\frac
{\alpha^2+\Lambda\beta^2}{(\tau\bar\tau)^2}.
\eea
\end{theorem}

{\bf Proof}: \;\; This is a direct consequence of \eqref{cpneat},
\eqref{rsym} and \eqref{rasym}.
\hfill $\Box$

To interpret the condition \eqref{Ncond} we
focus on the Lorentzian case, and recall that
$\Lambda$ is minus the cosmological constant in that case.
Clearly either $\alpha$ or $\beta$ have to vanish.
If $\beta=0$ then $\bm^2<0$ so $\bm$ has to be spacelike. If $\alpha=0$ then
\bea
\label{mcond}
\bm^2=-\frac{1}{\Lambda \beta^2},
\eea
 so $\bm$ is timelike if $\Lambda <0$ and spacelike if $\Lambda >0$.

As mentioned  in the introduction, the vector $\bm$ in appearing in the 
$r$-matrix \eqref{rmat1} is chosen to be timelike in the  usual formulation
of  the $\kappa$-Poincar\'e algebra and its de Sitter and anti-de Sitter analogues.
We also concentrate on this case from now on, 
and  therefore
set $\alpha=0$. We introduce a new constant $\kappa>0$
via
\bea
\label{kappadef}
\kappa= \sqrt{|\Lambda|}\beta,
\eea
which has the  dimension of mass.
 In the case  $\Lambda <0$  the condition \eqref{mcond}
on the timelike vector $\bm$ thus becomes
\bea
\bm^2= \frac{1}{{\kappa}^2}.
\eea

Note that the condition \eqref{match}, which matches the coefficients
in $K_\tau$ to the norm squared of the vector $\bm$ in $r_A$, is
a requirement  of the Fock-Rosly construction.  If we were only interested
in classifying Lie bialgebras we would have the weaker condition that $r_A$
satisfies the modified classical Yang-Baxter equation. However, since
the right-hand side of \eqref{rasym} is an invariant element of
$\gothh_\Lambda\otimes \gothh_\Lambda\otimes \gothh_\Lambda$ for any $\bm$,
this would impose no condition on $\bm$.

\section{Lie bialgebra structures  and their physical interpretation}
\label{bialsect}

In this section we are going to compute the Lie bialgebra structure associated 
to the $r$-matrix \eqref{rmatt}. We will  show  that we obtain
infinitesimal versions of the $\kappa$-Poincar\'e algebra and $\kappa$-Poincar\'e group, thus allowing us to interpret the  
generators of $\gothh_\Lambda$ and its dual physically.
We will see that the parameter $\kappa$ introduced in \eqref{kappadef} should 
indeed be identified with the parameter that gives the $\kappa$-Poincar\'e algebra
its name. We will also  study the Lie bialgebra structure in 
an alternative set of generators for the Lie algebra
$\gothh_\Lambda$ which will be useful for calculations in the next section.

We focus  initially on the 
case associated with $\kappa$-de Sitter symmetry, i.e. 
$\Lambda <0$ and  $\alpha=0$  so that $\tau=\theta\beta$; the limit 
$\Lambda \rightarrow 0$, which takes us from the de Sitter to the Poincar\'e
case, is discussed further below.
We introduce a  timelike vector $\bn$ which satisfies
\bea
\bn^2=-\Lambda.
\eea
Then we can meet the requirement \eqref{mcond}  by setting $\bm =
\frac{1}{\Lambda\beta}\bn$; the classical $r$-matrix  of
Theorem~\ref{rresult} becomes
\begin{align}
\label{rmat}
r=\frac{1}{\Lambda\beta}\left(\Lambda J_a\otimes J^a+P_a\otimes P^a
+\epsilon_{abc}n^a(J^b\otimes P^c-P^c\otimes J^b)\right).
\end{align}

This $r$-matrix defines a Lie bialgebra structure on $\gothh_\Lambda$.
We refer the reader to
\cite{CP} for a definition of a Lie bialgebra; for our
purposes the most important aspect of this structure is
that it gives rise to a Lie bracket on the dual space of
$\gothh_\Lambda$. To define this, one first computes
co-commutators from the $r$-matrix via
\begin{align}
\delta({J_a})&=(1\otimes \ad_{J_a} + \ad_{J_a}\otimes 1)r=\tfrac{1}{\Lambda\beta}(J_a\wedge  \bn\bP +P_a\wedge \bn\bJ),
\end{align}
where $X\wedge Y=X\otimes Y-Y\otimes X$.
Similarly
\begin{align}
\delta({P_a})&=(1\otimes \ad_{P_a} + \ad_{P_a}\otimes 1)r=\tfrac{1}{\Lambda \beta}\left(P_a\wedge  \bn\bP +\Lambda J_a\wedge \bn\bJ\
\right).
\end{align}
Note that the co-commutators only depend on the antisymmetric
part of $r$ since the symmetric part is a Casimir and hence invariant.
The Lie brackets $[\cdot,\cdot]_*$
in the dual of Lie algebra $\gothh_\Lambda$  are defined via
\bea
(\xi\otimes \eta)(\delta(X)) =  [\xi,\eta]_*(X),
\eea
where $X\in \gothh_\Lambda$,  $\xi,\eta \in \gothh_\Lambda^*$.
For the dual basis \eqref{dualbasis} the Lie
brackets are
\begin{align}
\label{dualcom}
 [P_a^*,P_b^*]_*&=\tfrac{1}{\Lambda\beta}
\left(P_a^*n_b-P_b^*n_a\right)\nonumber \\
[P_a^*,J_b^*]_*&=\tfrac{1}{\Lambda\beta}
\left(J_a^*n_b-J_b^*n_a\right)\nonumber \\
[J_a^*,J_b^*]_*&=\tfrac{1}{\beta}(P_a^*n_b-P_b^*n_a).
\end{align}
Interestingly we can recover these  from the bracket
\bea
\label{anfirst}
[P_a^*,P_b^*]_*=\tfrac{1}{\Lambda\beta}\left(P_a^*n_b-P_b^*n_a\right)
\eea
by setting $J^*_a=\theta P_a^*$.
Following \cite{we6} we denote the three-dimensional Lie algebra with
the  bracket \eqref{anfirst} by $\mathfrak{an}(2)$ .
Thus, just like the original Lie algebra $\gothh_\Lambda$ can be
obtained from  $\gothh$ by tensoring with the ring $R_\Lambda$, so the dual
$\gothh_\Lambda^*$  is obtained by tensoring  $\mathfrak{an}(2)$ with $R_\Lambda$.

In order to make contact with the usual formulation of the $\kappa$-Poincar\'e and $\kappa$-de Sitter
algebras and groups we pick $\bn=(\sqrt{|\Lambda|},0,0)$
and express the co-commutators and dual commutators in terms of the parameter $\kappa$
\eqref{kappadef}.  Then the co-commutators are 
\begin{align}
\label{kappaco}
\delta({J_a})=-\tfrac{1}{\kappa}(J_a\wedge P_0 +P_a\wedge J_0),\qquad
\delta({P_a})=-\tfrac{1}{\kappa}\left(P_a\wedge  P_0 + \Lambda J_a\wedge J_0\
\right).
\end{align}
and the dual brackets take the form
\begin{align}
\label{kappadualcom}
 [P_a^*,P_b^*]_*&=-\tfrac{1}{\kappa}
\left(P_a^*\delta^0_b-P_b^*\delta^0_a\right)\nonumber \\
[P_a^*,J_b^*]_*&=-\tfrac{1}{\kappa}
\left(J_a^*\delta^0_b-J_b^*\delta^0_a\right)\nonumber \\
[J_a^*,J_b^*]_*&=-\tfrac{\Lambda}{\kappa}\left(P_a^*\delta^0_b-P_b^*\delta^0_a\right).
\end{align}
These co-commutators and brackets  and their relation to the $\kappa$-de Sitter
algebra and group are discussed in \cite{BBH}. The letters for the generators 
used there, and in most 
of the literature on $\kappa$-Poincar\'e symmetries, differ from ours and 
are related to them as follows. The rotation generator $J_0$ is called 
$-J$, the boost generator $J_1$ is denoted $K_2$ and the boost generator $J_2$
is denoted $-K_1$. The notation for the  translation generators $P_a$  is the same
as ours, but their duals are denoted by $X_a$, to emphasise their interpretation 
as spacetime coordinates. In terms of $X_0=P^*_0$ and $X_j=P^*_j$,  with $j=1,2,$
the first of the brackets in \eqref{kappadualcom} gives the familiar space-time non-commutativity
\bea
\label{xcom}
[X_i,X_j]_*=0, \quad [X_0,X_j]_*=\tfrac{1}{\kappa} X_j, \qquad i,j=1,2.
\eea
The notation used in \cite{BBH} for $J^*_0$ is $\hat\theta$ to indicate its interpretation
as an angle, and the notation for $J^*_i$ is $\hat \xi_i$, $i=1,2$, to indicate the interpretation as
a boost parameter or rapidity. The interpretation suggested by this notation is thus in agreement with
our general remarks in Sect.~\ref{symreview}: the generators $P_0,P_1,P_2$ and $J, K_1,K_2$ are generators of the 
symmetry (Sklyanin) Poisson-Lie group, or equivalently  ``infinitesimal coordinates'' (coordinates near the identity)
of the  phase space (dual)  Poisson-Lie group. The  dual generators $X_0,X_1,X_2$ and $\hat \theta, \hat\xi_1,\xi_2$
are generators of the dual Poisson-Lie groups or, equivalently, 
 ``infinitesimal coordinates'' on the  symmetry Poisson-Lie group, 
with their names alluding to the  second viewpoint.
Finally, note that the limit $\Lambda \rightarrow 0$, which takes us from the $\kappa$-de Sitter algebra and group 
to the $\kappa$-Poincar\'e algebra and group, has to be accompanied by $\beta \rightarrow \infty$
in such a way that $\kappa$ stays fixed. With the limit taken in that way, all co-commutators and commutators
given above have a  smooth limit.

For the calculations in the next  section it is
convenient to replace the generators $P_a$  by
\bea
S_a=P_a+\epsilon_{abc}n^bJ^c, \quad \bn^2=-\Lambda.
\eea
The Lie brackets \eqref{liebra}  on $\gothh_\Lambda$
now take the form
\begin{align}
\label{liebr}
[J_a,J_b]=\epsilon_{abc}J^c,\quad[J_a,S_b]=\epsilon_{abc}S^c+n_bJ_a-\eta_{ab}\bn\bJ, \quad[S_a,S_b]=n_aS_b-n_bS_a.
\end{align}
The two Casimirs are
\begin{align}
K_t&=S_a\otimes J^a+J_a\otimes S^a\nonumber \\
K_s&=\frac{1}{\Lambda}\left(S_a\otimes  S^a-(\bn\bJ)
\otimes (\bn\bJ)+n_a\epsilon^{abc} (S_b\otimes J_c+ J_c\otimes S_b\right).
\end{align}
and the $r$-matrix \eqref{rmat} reads
\begin{align}
\label{rmattt}
r=\frac{1}{\Lambda\beta}\left(S_a\otimes S^a-(\bn\bJ)\otimes
(\bn\bJ)+2n_a\epsilon^{abc} S_b\otimes J_c\right).\end{align}

One advantage of the generators $S_a$ is that they generate
a subalgebra of $\gothh_\Lambda$ which is isomorphic to the Lie algebra
$\mathfrak{an}(2)$ encountered earlier \eqref{anfirst}. Explicitly, in
  the case $\Lambda<0$ and with the choice $\bn=(\sqrt{|\Lambda|},0,0)$
 one finds
\begin{align}
\label{ancoms}
[S_0,S_i]=\sqrt{|\Lambda|}S_i\quad [S_i,S_j]=0\qquad i,j=1,2.
\end{align}
It is worth commenting on the different role of the $\mathfrak{an}(2)$ bracket
here and in \eqref{xcom}. The brackets \eqref{ancoms} are part of the Lie algebra 
of $\gothh_\Lambda$ and 
 show  that the non-commutativity of 
infinitesimal  translations in the  corresponding
model spacetime is controlled by the inverse length 
scale $\sqrt{|\Lambda|}$. The brackets \eqref{xcom}, by contrast, are part of the 
dual Lie algebra  $\gothh_\Lambda^*$ and show that the non-commutativity of infinitesimal
translation in momentum space is controlled by  the inverse mass scale $1/\kappa$.

If we use the   basis
\bea
\label{basiss}
\tilde B= \{J_0,J_1,J_2,S_0,S_1,S_2\}
\eea
 instead of \eqref{basis}
we have a new dual basis
\bea
\label{dualbasiss}
\tilde B^*= \{\tJ_0^*,\tJ_1^*,\tJ_2^*,S_0^*,S_1^*,S_2^*\},
\eea 
which is related to the
dual basis \eqref{dualbasis} via
\bea
\label{dualbases}
S_a^*=P_a^*, \qquad \tJ_a^*= J_a^* +\epsilon_{abc}n^b(P^*)^c.
\eea
The dual brackets  now take the form
\begin{align}
\label{dualcoms}
 [S_a^*,S_b^*]_*&=\tfrac{1}{\Lambda\beta}
(S_a^*n_b-S_b^*n_a)\nonumber \\
[\tJ_a^*,S_b^*]_*&=\tfrac{1}{\Lambda\beta}
\left(\tJ_a^*n_b-\tJ_b^*n_a+n_a(\bn\times \bS^*)_b\right)\nonumber \\
[\tJ_a^*,\tJ_b^*]_*&=\tfrac{1}{\Lambda\beta}\left((\bn\times \btJ^*)_a^*n_b-(\bn\times \btJ^*)_b^*n_a\right)=-\tfrac{1}{\Lambda\beta}
\epsilon_{abc}(\Lambda \tJ^c+(\bn\btJ^*)n^c).
\end{align}

In the next section  we require the Lie brackets of $\gothh_\Lambda$ in yet a different basis, namely the
basis obtained by applying the isomorphism $\isom:\gothh_\Lambda^*
\rightarrow \gothh_\Lambda$ defined
in \eqref{assdef} to the basis $\tilde B^*$ \eqref{dualbasiss}
With $\tau=\theta\beta$, the map $\isom$ is simple in the  dual basis \eqref{dualbasis}
\bea
\isom(J^*_a)=\tfrac{1}{\beta}J_a,\qquad\isom(P^*_a)=\tfrac{1}{\Lambda\beta}
P_a.
\eea
Using the bijection \eqref{dualbases} one finds
\begin{align}
\isom(\tJ_a^*)=\tfrac{1}{\Lambda\beta}\left( -n_a \bn\bJ +\epsilon_{abc} n^b
S^c\right)\quad \isom(S_a^*)= \tfrac{1}{\Lambda\beta}\left(
S_a-\epsilon_{abc}n^b J^c\right),
\end{align}
and in terms of the generators
\bea
I_a=\isom(\tJ^*_a),\qquad R_a=\isom(S^*_a),
\eea
the Lie bracket takes the form
\begin{align}
\label{ddbr}
&[I_a,I_b]=
\tfrac{1}{\Lambda\beta}\epsilon_{abc}\left( \bn\bI n^c
+\Lambda I^c\right)\nonumber \\
&[R_a,R_b]=\tfrac{1}{\Lambda\beta}\left( \epsilon_{abc} I^c + n_b
R_a-n_aR_b\right)\nonumber \\
&[I_a,R_b]=\tfrac{1}{\Lambda\beta}\left( \eta_{ab} \bn \bI-n_b I_a
+n_a\epsilon_{bcd}n^c R^d-\epsilon_{abc}
n^c\bn\bR\right).
\end{align}

\section{Poisson-Lie structures }

The $r$-matrix discussed in Sect.~\ref{rsect} gives rise to Poisson
structures on the local isometry  groups $H$ in Table~1: the Sklyanin Poisson
structure  on $H$ and the dual Poisson structure. 
As explained in the introductory Section~\ref{symreview}, $H$  equipped 
with the Sklyanin bracket is a Poisson-Lie group which generalises the notion
of a symmetry group; it is the classical limit of the 
$\kappa$-Poincar\'e group (or its de Sitter or anti-de Sitter versions).
 On the other hand, the dual Poisson bracket on $H$
is the pull-back of the Poisson  structure on the dual Poisson-Lie group,
which generalises the notion of a particle phase space and is the classical
limit of the $\kappa$-Poincar\'e algebra (or its de Sitter or anti-de Sitter versions).

We now compute and
discuss the Sklyanin and dual Poisson brackets, focussing on the de Sitter
case and the situation where the  special vector appearing in the
$r$-matrix is timelike.  Since the expressions we obtain  are  complicated   we 
 discuss various approximations which give additional insights. In particular
we discuss their linearisations and relate them to the Lie bialgebras 
of the previous
section.

\subsection{Parametrisation and coordinates}

In order to compute Poisson structures on $H$ 
we need to pick coordinates on the group manifold. As explained in
\cite{we6} one obtains a convenient and unified description of the
local isometry groups by thinking of them as unit (pseudo)
quaternions over the ring $R_\Lambda$. To appreciate this point of
view, recall the defining relation \bea \label{quat}
e_ae_b=-\eta_{ab} +\epsilon_{abc}e^c \eea
 for the unit imaginary (pseudo) quaternions $e_0,e_1,e_2$. If we now set
\bea \label{quatrep} J_a=\frac 1 2 e_a,\qquad S_a=\frac\theta 2 e_a
+ \frac 1 2 \epsilon_{abc}n^be^c, \quad \bn^2=-\Lambda, \eea
 we obtain the commutation
relations \eqref{liebr} as a consequence of \eqref{quat} and of 
$\theta^2=-\bn^2=\Lambda$. Specialising to the Lorentzian case with
$\Lambda <0$ one can cast the realisation \eqref{quatrep} in more
familiar from by using the representation
 \begin{align} \label{su11rep}
\rho(e_0) =\bpm i & 0 \\ 0 & -i \epm,
\quad \rho(e_1) =\bpm 0 & i \\ -i  & 0 \epm, \quad
\rho(e_2) =\bpm 0 & 1
\\ 1 & 0 \epm. \end{align}
Then $\rho(J_a)$ are
the  familiar generators of the Lie algebra $\mathfrak{su}(1,1)$.
Since  $\Lambda <0$ we can identify $\theta $ with
$i\sqrt{|\Lambda|}$ and pick $\bn=(\sqrt{|\Lambda|},0,0)$. Then
\begin{align}
\rho(S_0)=\sqrt{|\Lambda|}\bpm -1 & 0 \\ 0 & 1 \epm,\quad
\rho(S_1)=\sqrt{|\Lambda|}\bpm 0 & 0 \\ 1 & 0 \epm,\quad
\rho(S_2)=\sqrt{|\Lambda|}\bpm 0 & 0 \\ -i & 0 \epm,
\end{align}
which are familiar generators of $\mathfrak{an}(2)$ with the brackets
\eqref{ancoms}.

According to Table~1,  the local isometry group
in the Lorentzian case with $\Lambda <0$
is  $SL(2,\CC)$.
We will use a parametrisation of this group which relies
on its factorisation into
an element $u\in SU(1,1)$ and an element $s\in AN(2)$.
As discussed in \cite{we6} such a factorisation is not
possible globally; however, the coordinates we obtain cover
``half'' the group manifold, including  a neighbourhood of the identity.
The idea is to
parametrise $SU(1,1)$ group elements via
 \begin{align}
\label{pdef} u=p_3+p^aJ_a,\quad  \frac{\bp^2}{4} +p_3^2 =1,
\end{align}
and $AN(2)$ elements via
\begin{align}
\label{qdef}
s=q_3+q^aS_a, \quad  q_3=\sqrt{1+(\bq\bn)^2/4}.
\end{align}
Here  $J_a$ and $S_a$ are as defined in \eqref{quatrep}; one can
make contact with familiar matrix representations of $SU(1,1)$ and $AN(2)$
by using the representations \eqref{su11rep} but this is not essential
in what follows.
We factorise an element $h\in SL(2,\CC)$ into an $SU(1,1)$ and an $AN(2)$
element parametrised as above:
\bea
h=\left(p_3+p^b J_b\right)\cdot
\left(q_3+q^bS_b\right).
\eea
Then we use the coordinate functions
\begin{align}
\label{pqdef}
 p^a:\; h \mapsto
p^a, \qquad
q^a:\; h \mapsto q^a.
\end{align}

For the physical interpretation of the coordinates $\{p^0,p^1,p^2\}$
it is important to recall that $SU(1,1)$ is the double cover
of the orthochronous Lorentz group $SO^+(2,1)$ in three dimensions.
One obtains the    $SO^+(2,1)$ matrix
$\Lambda^a_{\;\;b}$  associated to an $SU(1,1)$ element via the
 adjoint representation:
\begin{align}
\label{reprel} \left(p_3+p^aJ_a\right) v^a J_a\left( p_3+p^a
J_a\right)^{-1}=\left(p_3+p^aJ_a\right) v^a J_a\left( p_3-p^a
J_a\right)=\Lambda^a_{\;\;b}(\bp) v^b J_a
\end{align}
Using the relations \eqref{quat}
we  calculate the matrix elements $\Lambda^a_{\;\;b}(\bp)$ in \eqref{reprel} and
find
\begin{align}
\label{lambdamat} \Lambda^a_{\;\;b}(\bp)=
\left(1-\tfrac{1}{2}\bp^2\right)\delta^a_b +\tfrac{1}{2}p^ap_b
-p_3\epsilon^a_{\;\;bc}p^c\qquad p_3=\sqrt{1-\bp^2/4}.
\end{align}
For $\Lambda<0$, $\bn=(\sqcc, 0,0)$, we have the following 
expressions for the matrix elements  $\Lambda^a_{\;\;b}(\bp)$:
\begin{align}
\label{lambdaels}
&\Lambda^0_{\;\;0}(\bp)=1+\tfrac{1}{2}(p_1^2+p_2^2) &
&\Lambda^1_{\;\;1}(\bp)=1-\tfrac{1}{2}(p_0^2-p_2^2) &
&\Lambda^2_{\;\;2}(\bp)=1-\tfrac{1}{2}(p_0^2-p_1^2)\nonumber\\
&\Lambda^0_{\;\;i}(\bp)=\tfrac{1}{2} p^0p_i-p_3
\epsilon^0_{\;\;ik}p^k & &\Lambda^i_{\;\;0}(\bp)=\tfrac{1}{2}
p_0p^i-p_3 \epsilon^i_{\;\;0k}p^k &
&\Lambda^i_{\;\;j}(\bp)=\tfrac{1}{2} p^ip_j-p_3
\epsilon^i_{\;\;j0}p^0\\
&i\neq j,\quad i,j=1,2.\nonumber
\end{align}
The Poisson brackets we are going to compute in
this section are defined in terms of vector fields
on the group $H=SL(2,\CC)$. For the computation we need to pick
generators
 $X_\alpha$, $\alpha =1,\ldots,\text{dim}(\gothh_\Lambda)$, of
$\gothh_\Lambda$ and to
compute the right- and left-invariant vector
fields  $X_\alpha^L$ and $X_\alpha^R$  associated to the generators via
\begin{align}
\label{lrvecfields} X_\alpha^Lf(h)=\frac{d}{dt}|_{t=0}
f(e^{-tX_\alpha}h),\qquad X_\alpha^Rf(h)=\frac{d}{dt}|_{t=0} f(h
e^{tX_\alpha})\qquad\forall h\in H, f\in\cif(H).
\end{align}
Expanding the classical
 $r$-matrix in terms of these generators
 $r=r^{\alpha\beta}X_\alpha\otimes X_\beta$, the Sklyanin Poisson-Lie structure
is defined via the bi-vector 
\begin{align}
&B_S=\tfrac{1}{2}r^{\alpha\beta}\left(X_\alpha^R\wedge X_\beta^R- X_\alpha^L\wedge
X_\beta^L\right).\label{skylb}
\end{align}
The notion of a Poisson bi-vector is discussed in \cite{CP}, but for our purpose
it is sufficient to know that 
the bracket of two functions $f,g\in\cif(H)$ is obtained by
acting with the vector fields on the functions $f$ and $g$ as spelled  out in
\eqref{lrvecfields}.
The second Poisson structure on $H$ which we  want to compute
is the dual Poisson structure discussed in the introduction. Its
Poisson bi-vector is
\begin{align}
B_D=\tfrac{1}{4} r^{\alpha\beta}\left(X_\alpha^R\wedge X_\beta^R+ X_\alpha^L\wedge
X_\beta^L\right)+\tfrac{1}{2}r^{\alpha\beta} X_\alpha^R\wedge X_\beta^L\label{dualb}.
\end{align}

For our computations we are going to use the basis $\tilde B$
\eqref{basiss} of $\gothh_\Lambda$. The expressions for the left-and
right-invariant vector fields generated by $J_a$ and  $S_a$ in terms
of the coordinate functions $p^a, q^a$  were first given in 
\cite{we6}. We reproduce them here for the convenience of the reader:
\begin{align}
&{J_a^L}
p^b=-\eta^{ab}p_3+\tfrac{1}{2}\epsilon^{abc}p_c\qquad J_a^Rp^b=\frac{k_-(\bq)}{k_+(\bq)}\left(
p_3\delta_a^b+\tfrac{1}{2}\epsilon_{a}^{\;\;bc}p_c\right)+\frac{q_a}{k_+(\bq)}\left(
p_3n^b+\tfrac{1}{2}\epsilon^{bcd}p_cn_d\right)\nonumber\\
&J_a^L q^b=0 \qquad \qquad\qquad\qquad\;\; J_a^Rq^b=\frac{1}{k_+(\bq)}\left(q_3\epsilon_a^{\;\;bc}q_c-\tfrac{1}{2}q^a\epsilon^{bcd}n_cq_d\right)\label{jrq}\\
\nonumber\\
&{S_a^R} q^b=k_+(\bq)\,\eta^{ab}-\tfrac{1}{2} n^a
q^b\quad\;\; \; S_a^L q^b=-\Lambda^a_{\;\;c}(\bp)\left(\left(q_3-\tfrac{1}{2}\bq\bn\right)\eta^{bc}+\tfrac{1}{2}n^cq^b\right)
\\
&S_a^Rp^b=0\qquad \qquad\qquad\qquad\;\;\; S_a^Lp^b=\tfrac{p_3}{2}p^b\epsilon_{acd}p^cn^d+p_3^2
n_ap^b+\tfrac{1}{4}\bp\bn p_ap^b-p_an^b,\nonumber\end{align}
where $k_\pm(\bq)=q_3\pm \tfrac{1}{2}\bq\bn$ and $\Lambda^a_{\;\;c}(\bp)$ is given by \eqref{lambdamat} and \eqref{lambdaels}.
The evaluation of the
brackets \eqref{skylb} and \eqref{dualb} is straightforward but lengthy and relies on the repeated use of the identity
\begin{align}
\label{helpidentity}
x^b\epsilon^{acd}y_cx_d-x^a\epsilon^{bcd}y_cx_d=\epsilon^{abc}(\bx\by
x_c-\bx^2 y_c).
\end{align}
We therefore present only its result.

\subsection{The Sklyanin bracket and the $\kappa$-Poincar\'e group}

The Poisson-Lie group $SL(2,\CC)$ equipped with the Sklyanin bracket is the 
classical limit of the $\kappa$-de Sitter group. 
 In terms of the coordinates $p^a,q^a$ the Sklyanin bracket for the $r$-matrix \eqref{rmattt}
\eqref{skylb} takes the form
\begin{align}
&\{p^a,p^b\}=-\tfrac{1}{\Lambda\beta} p_3\epsilon^{abc}\left( \bp\bn n_c+\Lambda p_c\right)\nonumber\\
&\{q^a,q^b\}=\tfrac{1}{\Lambda\beta} q_3\left(n^bq^a-n^aq^b\right)\nonumber\\
&\{p^a,q^b\}=\tfrac{1}{\Lambda\beta}(q_3-\tfrac{1}{2}\bn\bq)\left(
\tfrac{p_3}{2}p^a\epsilon^{bcd}n_cp_d -n^ap^b +p_3^2
p^an^b+\tfrac{1}{4}\bp\bn p^ap^b\right)\nonumber\\
&\qquad\quad\;\;+\tfrac{1}{\Lambda\beta}\left( -\tfrac{1}{2}\bp\bn q^bn^a
-\tfrac{\Lambda}{2}p^aq^b(1-\bp_\bot^2/4)+p_3n^a\epsilon^{bcd}n_cq_d-\tfrac{1}{2}\epsilon^{akl}p_kn_l\epsilon^{bcd}q_cn_d\right),\label{skylbco}
\end{align}
where $\bp^2_\bot=\tfrac{1}{\Lambda} (\bp^2+(\bp\bn)^2)$.

To understand the structure of these brackets we first consider
their linearisations near the identity:
\begin{align}
&\{p^a,p^b\}=-\tfrac{1}{\Lambda\beta} \epsilon^{abc}\left( \bp\bn n_c+\Lambda p_c\right)+\mathcal O(\bp^2)\nonumber \\
&\{q^a,q^b\}=\tfrac{1}{\Lambda\beta} \left(n^bq^a-n^aq^b\right)+\mathcal O(\bq^2) \nonumber \\
&\{p^a,q^b\}=\tfrac{1}{\Lambda\beta}\left(n^b p^a-n^ap^b
+n^a\epsilon^{bcd}n_cq_d\right)+\mathcal O
(\bp^2,\bq^2,\bp\bq).\label{linskyl}
\end{align}
This agrees  with expression  \eqref{dualcoms} for the dual Lie bracket,
as it should according to the general remarks in Sect.~\ref{symreview}:
the dual Lie bracket is the
infinitesimal version of the Sklyanin Poisson bracket.

Next we consider the limit  $\Lambda\rightarrow 0$. 
As for the dual Lie brackets in the previous section, we take the limit 
$\beta \rightarrow \infty$ at the same time
 in such a way that  $\kappa$, as  defined
in \eqref{kappadef},  remains constant.
Recalling  that $\Lambda<0$ and taking $\bn=(\sqrt{|\Lambda|},0,0)$, the brackets
\eqref{skylbco} become, in this limit,
\begin{align}
&\{p^a,p^b\}=0\nonumber\\
&\{q^a,q^b\}=-\tfrac{1}{\kappa}\left(\delta ^b_0 q^a-\delta^a_0 q^b\right)\nonumber\\
&\{p^a,q^b\}=-\tfrac{1}{\kappa} \left(
\tfrac{p_3}{2}p^a\epsilon^{b0d}p_d -\delta^a_0 p^b +p_3^2 \delta^b_0
p^a+\tfrac{p_0}{4}
p^a p^b\right)\label{pqbrack3}
\end{align}
We thus find that the coordinates $q^a$ have the $\mathfrak{an}(2)$
brackets \eqref{ancoms}, while the bracket of the $p^a$ vanishes and
the bracket of $q^a$ and $p^a$ yields a term proportional to
$p^a$. In order to relate these Poisson brackets to the commutators of the 
$\kappa$-Poincar\'e group  in its usual formulation
\cite{Zakrzewski,KM},  we work with the expressions
\eqref{lambdaels}
 for the $SO^+(2,1)$ matrix associated to the
coordinates $p^a$. Using expression 
\eqref{pqbrack3} for the bracket in order $\mathcal O(\sqcc)$, we can
calculate the brackets of the matrix elements
$\Lambda^a_{\;\;b}$ with the position coordinates $q^c$
explicitly and obtain 
\begin{align}
&\{\Lambda^a_{\;\;b}, \Lambda^c_{\;\;d}\}=0\nonumber\\
&\{q^a, q^b\}=-\tfrac{1}{\kappa}\left(\delta ^b_0 q^a-\delta^a_0
q^b\right)\nonumber\\
&\{\Lambda^a_{\;\;b}, q^c\}=-\tfrac{1}{\kappa}\left(
(\Lambda^a_{\;\;0}-\delta^a_{\;\;0})\Lambda^c_{\;\;b}
+\eta^{ac}(\Lambda^0_{\;\;b}-\delta^0_{\;\;b})\right).
\end{align}
As expected, this is the classical limit of the $\kappa$-Poincar\'e group as given in \cite{Zakrzewski,KM}.

\subsection{The dual Poisson  structure}
\label{dualsec}

Continuing with the de Sitter case ( $\Lambda <0$   and $H=SL(2,\CC)$)
we now compute the dual Poisson structure \eqref{dualb}   on $H=SL(2,\CC)$,
using the $r$-matrix \eqref{rmattt}. As explained in the introduction, this is the pull-back of 
the Poisson structure on the Poisson-Lie  group dual to the Sklyanin Poisson-Lie group. 
Concretely, the pull-back means that the dual Poisson-Lie group  is coordinatised in terms
of the original group $SL(2,\CC)$, and its Poisson brackets are given in terms of the 
coordinates \eqref{pqdef} on $SL(2,\CC)$. We compute the dual Poisson structure in 
this way because this is  how it arises in the Fock-Rosly
 construction: for each puncture, the  auxiliary phase space of the Fock-Rosly construction
contains a copy of the group $H=SL(2,\CC)$,  equipped with the dual Poisson bracket.
Since punctures are interpreted as particles, the dual Poisson bracket  thus gives the 
Poisson structure of the particle phase space - in agreement with our general remarks
about the interpretation of the dual Poisson-Lie group as a generalised phase space.

The existence of the  pull-back, which is assumed in the general formula \eqref{dualb}
for the dual bracket, depends on the  non-degeneracy of the  symmetric part of the   $r$-matrix
\eqref{rmattt}.  In the Fock-Rosly construction that symmetric part is equal to the Casimir $K_\tau$, 
which is non-degenerate by assumption. However, in the limit $\Lambda\rightarrow 0$, the Casimir $K_\tau$
(with $\alpha=0$) becomes singular. This leads to singularities in the dual Poisson  brackets,
as we shall see below.

The dependence of the  dual bracket on the symmetric part of $r$-matrix is manifest in the formula \eqref{dualb}.
This should be contrasted with the Sklyanin bracket,
which only depends on the anti-symmetric part. Explicitly, we find
the following expression for \eqref{dualb}  in terms of the
coordinates \eqref{pqdef}:
\begin{align}
\label{pbrack1} &\{p^a,p^b\}=\tfrac{1}{\beta\Lambda} p_3\epsilon^{abc}\left( \bp\bn n_c+\Lambda p_c\right)\\
&\{q^a,q^b\}=\tfrac{1}{\beta\Lambda}\left(p_3\epsilon^{abc}p_c+ q_3p^2_3(q^an^b-q^bn^a)\right)\nonumber \\
&\qquad\quad +\tfrac{1}{\beta\Lambda}\left(
\tfrac{q_3}{4} \bp\bn(q^ap^b-p^aq^b)+ \tfrac{p_3}{4}\bq\bn(q^b\epsilon^{acd}n_cp_d-q^a\epsilon^{bcd}n_cp_d)\right)\label{qqtrouble}\nonumber\\
&\{p^a,q^b\}=\tfrac{1}{\beta\Lambda}\left( p_3 n^b\epsilon^{acd} n_cq_d+\tfrac{p_3}{4}\bq\bn p^a\epsilon^{bcd} n_cp_d +\Lambda p_3\epsilon^{abc} q_c +q_3\bp\bn \eta^{ab} -p_3^2q_3 n^bp^a -\tfrac{q_3}{4}\bp\bn p^ap^b\right)\nonumber
\end{align} with
$p_3,q_3$ given by \eqref{pdef} and \eqref{qdef}.

The functions
$\bp\bq$ and  $p_3q_3-\tfrac{1}{4}\epsilon_{abc}n^ap^bq^c$
are Casimir functions of this Poisson bracket. As explained in
\cite{we6} these functions have a simple geometrical interpretation
in terms of the quaternionic language briefly introduced above.
In  particular, they are constant on conjugacy classes
of the group $H=SL(2,\CC)$, which, according to the general
theory of dressing transformations, are precisely the symplectic leaves
of the dual Poisson structure.


 The linearisation 
of the brackets \eqref{pbrack1}  near the identity is given by
\begin{align}
&\{p^a,p^b\}=\tfrac{1}{\beta\Lambda} \epsilon^{abc}\left( \bp\bn\, n_c+\Lambda
p_c\right)+\mathcal O (\bp^2)\\
&\{q^a, q^b\}=\tfrac{1}{\beta\Lambda}\left( \epsilon^{abc} p_c +q^an^b-n^a
q^b\right)+\mathcal O (\bp^2,\bp\bq, \bq^2)\nonumber\\
&\{p^a,q^b\}=\tfrac{1}{\beta\Lambda}\left( \bp\bn
\,\eta^{ab}-n^bp^a+n^a\epsilon^{bcd}n_cq_d -\bq\bn
\epsilon^{abc}n_c\right)+\mathcal O (\bp^2,\bp\bq, \bq^2).\nonumber
\end{align}
Comparing with \eqref{ddbr}
one finds  agreement with the Lie bracket on $\gothh_\Lambda$ when
expressed in terms of the images under the map $\isom$ of the
generators of the dual Lie algebra $\gothh_\Lambda^*$. 
This confirms that the dual Poisson  structure
reduces to  the Lie bracket of  $\gothh_\Lambda$ near the identity.
Note that the existence of the map  $\isom$   depends on
the non-degeneracy of the $\Ad$-invariant symmetric bilinear form $(\cdot, \cdot)_\tau$,
which is in turn equivalent to the non-degeneracy of the Casimir $K_\tau$.
Since the latter is assumed in the expression \eqref{dualb} for the dual Poisson
 structure, it is not surprising that  we need the identification between
$\gothh_\Lambda$ and $\gothh_\Lambda^*$ via  \eqref{assdef} here.

The dual Poisson structure does not have a well-defined limit
when $\Lambda\rightarrow 0$. Even when we take $\beta\rightarrow \infty$
in such a way that $\kappa$ \eqref{kappadef}  remains constant,
the first term in the second line of \eqref{qqtrouble} tends to infinity. As explained
at the beginning of this subsection, this singularity arises because
the pull-back of the Poisson structure on the dual Poisson-Lie group to
$SL(2,\CC)$ becomes singular in the limit $\Lambda \rightarrow 0$.
Thus, even though the  dual Poisson-Lie group does have a smooth limit as $\Lambda\rightarrow
0$, the pull-back of its Poisson structure to $SL(2,\CC)$ becomes ill-defined.
We discuss implications of this result in the conclusion.

\section{Conclusion and outlook}

In this paper we addressed the question if $\kappa$-Poincar\'e
symmetry and its de Sitter and anti-de Sitter analogues in three
dimensions can  be associated to Chern-Simons theories with,
respectively, the Poincar\'e, de Sitter and anti-de Sitter
group as gauge group.
In practice this meant checking if the  
classical  $r$-matrices obtained from  the $\kappa$-Poincar\'e algebra and its
de Sitter and anti-de Sitter analogues  are compatible 
with those Chern-Simons theories  via the Fock-Rosly construction.
We showed that,  if one insists on the 
vector appearing in the $r$-matrix being timelike, only 
 the  $\kappa$-de Sitter algebra can be  associated  to a   Chern-Simons
theory in this way. The relevant  Chern-Simons action  is based on  the 
 non-degenerate symmetric form $s(\cdot,\cdot)$, which  is 
not the one used in 3d gravity. 

The association between the $\kappa$-de Sitter algebra and Chern-Simons gauge theory opens up the possibility of constructing a multi-particle system with $\kappa$-de Sitter symmetry.  Given the compatibility between the  classical $r$-matrix for the $\kappa$-de Sitter algebra with the Chern-Simons action, one can  use the Fock-Rosly method to
construct the phase space for an arbitrary number of interacting particles coupled to Chern-Simons theory. 
The Poisson-structure on that phase
space is invariant under the Sklyanin Poisson-Lie group 
based on the same $r$-matrix.
In this paper
we computed the Sklyanin Poisson bracket and  the dual bracket.
 The Fock-Rosly bracket on the  $n$-particle phase can be 
 expressed in terms of $n$ copies of the dual bracket; doing this explicitly
and interpreting the resulting phase space is left
for  future work.

The limit $\Lambda\rightarrow 0$ is subtle, and deserves further comments.
On can recover
the $\kappa$-Poincar\'e group and algebra, as well as all 
its associated  Lie bialgebra
and Poisson-Lie group structures, from the corresponding de Sitter
version by taking 
the limit $\Lambda\rightarrow 0$ while keeping $\kappa$ fixed.
However, the  symmetric 
form $s(\cdot,\cdot)$  used in the  Chern-Simons theory associated to 
$\kappa$-de Sitter symmetry
degenerates in this limit, and, as a result, the basic Poisson brackets 
\eqref{eebr} become ill-defined. Not surprisingly, the Fock-Rosly construction
of an auxiliary phase space also 
fails in this situation. We saw this in our discussion of the dual Poisson-Lie
structure at the end of Sect.~\ref{dualsec}: even though
the dual Poisson-Lie group does have a smooth limit,  the pull-back
of its Poisson structure to the  group $H$ (which is required in the 
Fock-Rosly construction) does not. 
We therefore conclude that one cannot associate a Chern-Simons model to 
the  $\kappa$-Poincar\'e group and algebra in three dimensions
 by the method used in the present paper.

It is worth stressing that the requirement of compatibility of 
an $r$-matrix with a Chern-Simons theory is a much stronger 
requirement than the modified classical Yang-Baxter equation,
which is needed for the construction of Lie bialgebras
and Poisson-Lie groups from the same $r$-matrix. In particular,
the $r$-matrix
\eqref{rmat1} gives rise to Lie bialgebras and Poisson-Lie groups
for any choice
of the vector $\bm$.
The difference between these two requirements is related to the fact that  the 
limit $\Lambda\rightarrow 0$  works well for the Lie bialgebras and Poisson-Lie groups, but is ill-defined for the auxiliary phase space of the Fock-Rosly
construction.

If one insists on the 
vector appearing in the $r$-matrix being timelike
 it is impossible to associate
either the  $\kappa$-Poincar\'e algebra or its
de Sitter and anti-de Sitter analogues  to the Chern-Simons actions 
of 3d gravity, which are based on the symmetric  form
$t(\cdot, \cdot)$. This follows directly from the condition \eqref{Ncond}.
The $r$-matrices which are compatible with the Chern-Simons action of 3d 
gravity are discussed systematically in \cite{we6}. Their form is related 
to the fact that the Lie algebra $\gothh_\Lambda$ with the  non-degenerate
symmetric  form \eqref{pair} has the structure of a classical double.
The corresponding Hopf algebras  are all quantum doubles, and 
quite different from the $\kappa$-Poincar\'e, $\kappa$-de Sitter and
$\kappa$-anti-de Sitter
algebras, which all  have the structure of bicrossproducts \cite{MR,Majid}.

Claims in the literature   that  $\kappa$-Poincar\'e symmetry does arise
in 3d gravity tend to be   based on the algebra structure alone and ignore
the co-algebra. This is the case, for example in  \cite{FKGS}, which focuses on the 
algebra structure.  However,  as 
algebras  both the $\kappa$-Poincar\'e algebra 
 and the quantum double of the Lorentz group (which does arise in 3d gravity
\cite{we2})  are isomorphic. This is an immediate consequence of the fact  that both are isomorphic  to the universal enveloping
algebra of the Poincar\'e Lie algebra. For the $\kappa$-Poincar\'e algebra
this can be shown explicitly by writing it in a suitable basis \cite{KGN2},
and for the quantum double of the Lorentz group this is obvious in the usual
 formulation \cite{bamus, we2}. 
To distinguish the $\kappa$-Poincar\'e algebra 
 and the quantum double of the Lorentz group,
and to show that either arises in 3d gravity  one
 has to take the full Hopf algebra structure into account,
and this was not done in \cite{FKGS}. 
 The relation between  bicrossproducts on the one hand 
and the quantum  doubles arising in 3d gravity
is discussed further in \cite{sissatalk} and \cite{MS}. The upshot of the discussion there
and  in the current paper is that $\kappa$-Poincar\'e symmetry
is not directly related to 3d gravity.  It is possible to establish a connection
using the notion of  semi-duality \cite{MS}, but the physical significance of this
remains to be clarified.

\section*{Acknowledgements}
BJS thanks Jerzy Kowalski-Glikman for interesting him in the topic
of this paper. Both CM and BJS thank Jerzy Kowalski-Glikman and
Florian Girelli for extensive discussions and correspondence. This
research was supported by Perimeter Institute for Theoretical
Physics. Research at Perimeter Institute is supported by the
Government of Canada through Industry Canada and by the Province of
Ontario through the Ministry of Research \& Innovation.

\appendix

\section{Killing forms and a three-dimensional identity}

For any Lie algebra
the Killing form is  the $\Ad$-invariant symmetric bilinear form defined via
\bea
k(X,Y)=\tr (\ad(X)\ad(Y)).
\eea
Its matrix relative to a basis of the Lie algebra
 can be expressed entirely in terms of the structure constants
in that basis.
For  both the Lie algebras $\mathfrak{su}(2)$ and $\mathfrak{su}(1,1)$ with
the generators described in the main text,
 the matrix of the 
Killing form turns out to be
\bea
k(J_a,J_b)=-2\eta_{ab},
\eea
with $\eta=$diag$(1,1,1)$ in the Euclidean case
and $\eta=$diag$(1,-1,-1)$ in the Lorentzian case.

In the following we make use of two
identities
which hold for any semisimple Lie algebra, namely the Jacobi identity
\bea
[X,[Y,Z]]+[Y,[Z,X]]+[Z,[X,Y]]=0
\eea
and the invariance of the Killing form
\bea
k([X,Y],Z) + k(Y,[X,Z])=0.
\eea
However, we also need a special identity for the double commutator,
which holds only for the
semi-simple three dimensional Lie algebras $\mathfrak{su}(2)$ and
$\mathfrak{su}(1,1)$:
\bea
\label{special}
[X,[Y,Z]]=\eta(X,Z)\,\,Y -\eta(X,Y)\,\,Z.
\eea
This is equivalent to \eqref{helpidentity} and can be proved by checking it on a basis.

The following lemma can be proved by repeated but straightforward
 application of the
Jacobi identity, the invariance of the Killing form, and the special
identity \eqref{special}. 

\begin{lemma} Let $X,Y,Z,N$ be four elements of either $\mathfrak{su}(2)$
or $\mathfrak{su}(1,1)$. Then
\bea
\label{use2}
\eta([X,[Y,N]],[Z,N]) &+&
\eta([Y,[N,X]],[Z,N])+
\eta(Z,[[X,N],[Y,N]])\nonumber \\
&=&\eta(N,N)\eta([X,Y],Z)
\eea
\end{lemma}

The equation \eqref{use2} is equivalent to
the identity
\bea
\label{use4}
[[\,J_a\otimes[N,J^a],J_b\otimes [N,J^b]\,]]=
\eta(N,N)\,\,\epsilon_{abc}J^a\otimes J^b \otimes J^c,
\eea
as can be seen by taking the inner product with  $X\otimes Y\otimes Z$.

\end{document}